\documentclass[preprint,proceedings]{rmaa}

\suppressfulladdresses % by popular request

\usepackage{paralist}

\usepackage{psfrag,color}

\SetYear{2006}
\SetConfTitle{ }

\title{The mid-infrared emission of local Luminous Infrared Galaxies} 

\author{
  Almudena Alonso-Herrero\altaffilmark{1} 
}

\altaffiltext{1}{DAMIR, Instituto de Estructura de la Materia,
  CSIC, Madrid, Spain
  (aalonso@damir.iem.csic.es).} 

\shortauthor{Alonso-Herrero}
\shorttitle{Mid-IR emission of local LIRGs}

\listofauthors{Alonso-Herrero}
\indexauthor{Alonso-Herrero, A.}

\abstract{In this paper we use the complementary imaging capabilities of {\it Spitzer}
(sensitivity) and Gemini-South/T-ReCS (spatial resolution) to study the mid-IR
properties of local ($d < 75\,$Mpc) LIRGs. 
The T-ReCS $8-10\,\mu$m imaging observations of LIRGs have allowed us
to spatially resolve the nuclear emission (star formation and/or AGN) and that of
H\,{\sc ii} regions in the central $3-7\,$kpc regions of LIRGs. 
From the comparison of the 
$8\,\mu$m/Pa$\alpha$ ratios of the integrated vs. resolved H\,{\sc ii} regions
of LIRGs, we infer the existence of an $8\,\mu$m 
diffuse component, not directly related to the ionizing stars,
that can be as luminous as that from the resolved H\,{\sc ii} regions.  
We conclude that although the mid-IR integrated 
luminosity of galaxies undergoing dusty, intense star formation is a good indicator
of the star formation rate (SFR), the empirical calibrations should be based on the 
integrated emission of nearby galaxies, not that of H\,{\sc ii} regions
alone. To this end we provide a calibration of the SFR in terms of the integrated 
$24\,\mu$m luminosity that can be used for distant dusty galaxies.}

\resumen{En este art\'{\i}culo se combinan la sensibilidad de {\it Spitzer}
y la alta resoluci\'on espacial de T-ReCS en el Gemini-South para estudiar
las propiedades en el infrarrojo (IR) medio de galaxias
luminosas infrarrojas (LIRGs) locales ($d < 75\,$Mpc). Las im\'agenes de
T-ReCS en $8-10\,\mu$m nos han permitido separar la emisi\'on nuclear
(formaci\'on estelar y/o AGN) de la de las regiones H\,{\sc ii}
en las zonas centrales ($3-7\,$kpc). 
De la comparaci\'on de los cocientes  $8\,\mu$m/Pa$\alpha$ integrados y de
regiones H\,{\sc ii} resueltas se deduce la existencia de 
una componente de emisi\'on difusa en $8\,\mu$m. Esta emisi\'on difusa no
est\'a directamente relacionada con las poblaciones estelares ionizantes, y 
puede ser tan luminosa como la que proviene de regiones H\,{\sc ii}.
Por lo tanto, aunque la emisi\'on en el IR medio es un buen indicador de la
tasa de formaci\'on estelar (SFR) de galaxias con polvo, las calibraciones han
de hacerse usando las propiedades integradas de galaxias, y no las
de regiones H\,{\sc ii} \'unicamente.  
Se presenta una calibraci\'on de la SFR en
funci\'on de la luminosidad en $24\,\mu$m que es aplicable a galaxias lejanas 
con polvo.}

\addkeyword{Galaxies: Infrared}
\addkeyword{Galaxies: Star Formation}
\addkeyword{Galaxies: Seyfert}
\addkeyword{H~II regions}

\begin{document}
% Typeset article header
\maketitle

\section{Introduction}
\label{sec:intro}

The importance of infrared (IR) bright galaxies 
has been increasingly appreciated since their discovery more than 30 years ago
(Rieke  \& Low 1972), and the detection of large numbers  by  {\it IRAS}. 
Although the Ultraluminous IR Galaxies (ULIRGs, $L_{\rm IR} > 
10^{12}\,{\rm   L}_\odot$)
get much of the attention because they are so dramatic,
Luminous IR Galaxies (LIRGs --- $L_{\rm IR} = 10^{11}$ to $10^{12}\,{\rm
  L}_\odot$) are much more common, accounting for $\sim$ 5\% of the local
IR background compared with $<$ 1\% for the ULIRGs (Lagache et al. 2005). 
LIRGs take part in the controversy over
the formation of AGN, and its relation to star formation and high
levels of IR emission (see review by Sanders \& Mirabel 1996), and 
should also include former ULIRGs where the SF is dying off. 
Deep {\it Spitzer} detections at $24\,\mu$m are dominated by LIRGs (Le Floc'h et al. 
2005; P\'erez-Gonz\'alez et al. 2005), and LIRGs contribute nearly 50\% of the cosmic
IR background at  $ z \sim 1$  (Lagache et al. 2005).
Interpretation of the {\it Spitzer} observations at high-$z$ therefore
depends on a deep understanding of LIRGs in the local universe.

With the advent of a new generation of mid-IR instruments with superior sensitivity and 
spatial resolution we can now investigate whether the
mid-IR emission is a good  indicator of the star formation rate (SFR) in dusty
galaxies. The mid-IR emission has the advantage of not being affected by 
the contribution from cold dust heated by old stars that
  may dominate the far-IR luminosities. {\it Spitzer} observations of nearby {\it normal} 
galaxies show similar $8-24\,\mu$m
and hydrogen recombination line morphologies (e.g., Helou et al. 
2004; Hinz et al. 2004; Calzetti et al. 2005, CAL05 hereafter;
P\'erez-Gonz\'alez et al. 2006, PGPG06 hereafter; Alonso-Herrero et
al. 2006b). At longer wavelengths ($160\,\mu$m) the individual H\,{\sc ii}
regions (current star formation) contribute much less than at $24$ and
$70\,\mu$m (Hinz et al. 2004).
From a quantitative point of view there is a good correlation 
between the Pa$\alpha$ or H$\alpha$ luminosity (corrected for extinction) and the
$24\,\mu$m luminosity of resolved  H\,{\sc ii} 
regions (CAL05; PGPG06). However there are galaxy-to-galaxy
  variations in the $24\,\mu$m luminosity to SFR ratios of UV-selected
  starbursts and ULIRGs (CAL05). CAL05 and PGPG06, on the other hand,
  questioned the use of the IRAC $8\,\mu$m emission as a SFR tracer, whereas
  Wu et al. (2005) find a good correlation between the $8\,\mu$m (and
  $24\,\mu$m) and H$\alpha$
  emission of local star forming galaxies in the {\it Spitzer} First Look Survey.

We present a detailed study of the mid-IR properties of 
local LIRGs  taking advantage of both the 
sensitivity of {\it Spitzer} imaging observations and the high spatial
resolution of the  Thermal-Region Camera Spectrograph 
(T-ReCS; Telesco et al. 1998) 
on the 8.1\,m Gemini-South Telescope. In particular we investigate the issue
of whether the mid-IR emission is a good tracer of the SFR in dusty galaxies.

\begin{figure*}[!t]\centering
\includegraphics[angle=-90,width=2.\columnwidth]{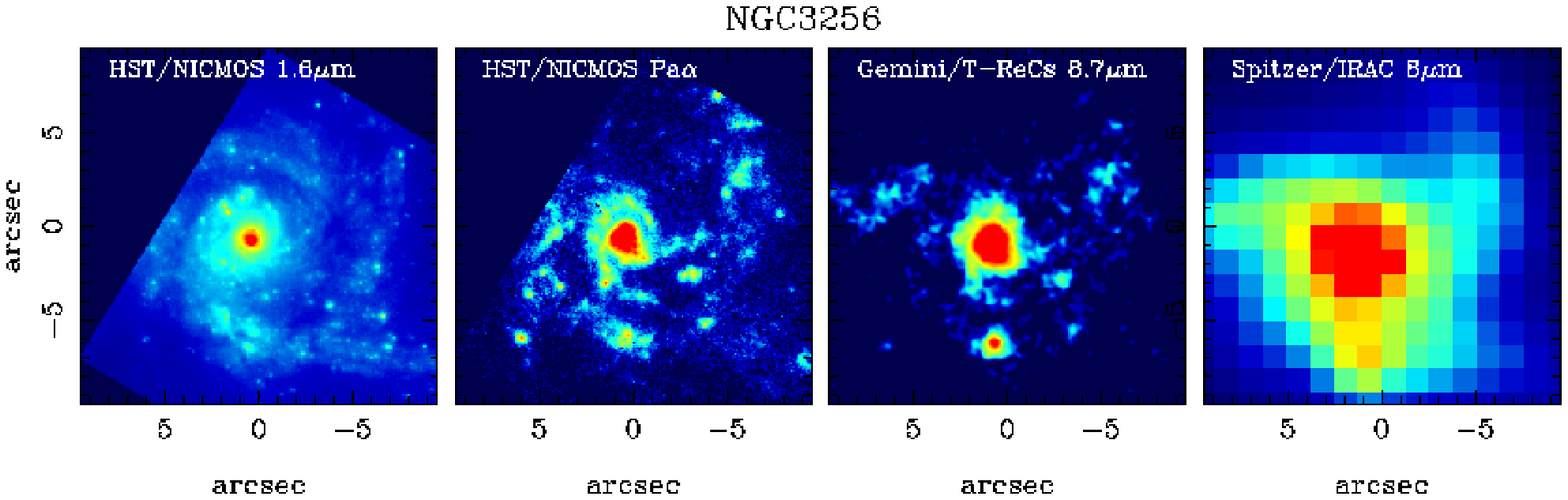}

\vspace{0.5cm}

\includegraphics[angle=-90,width=2.\columnwidth]{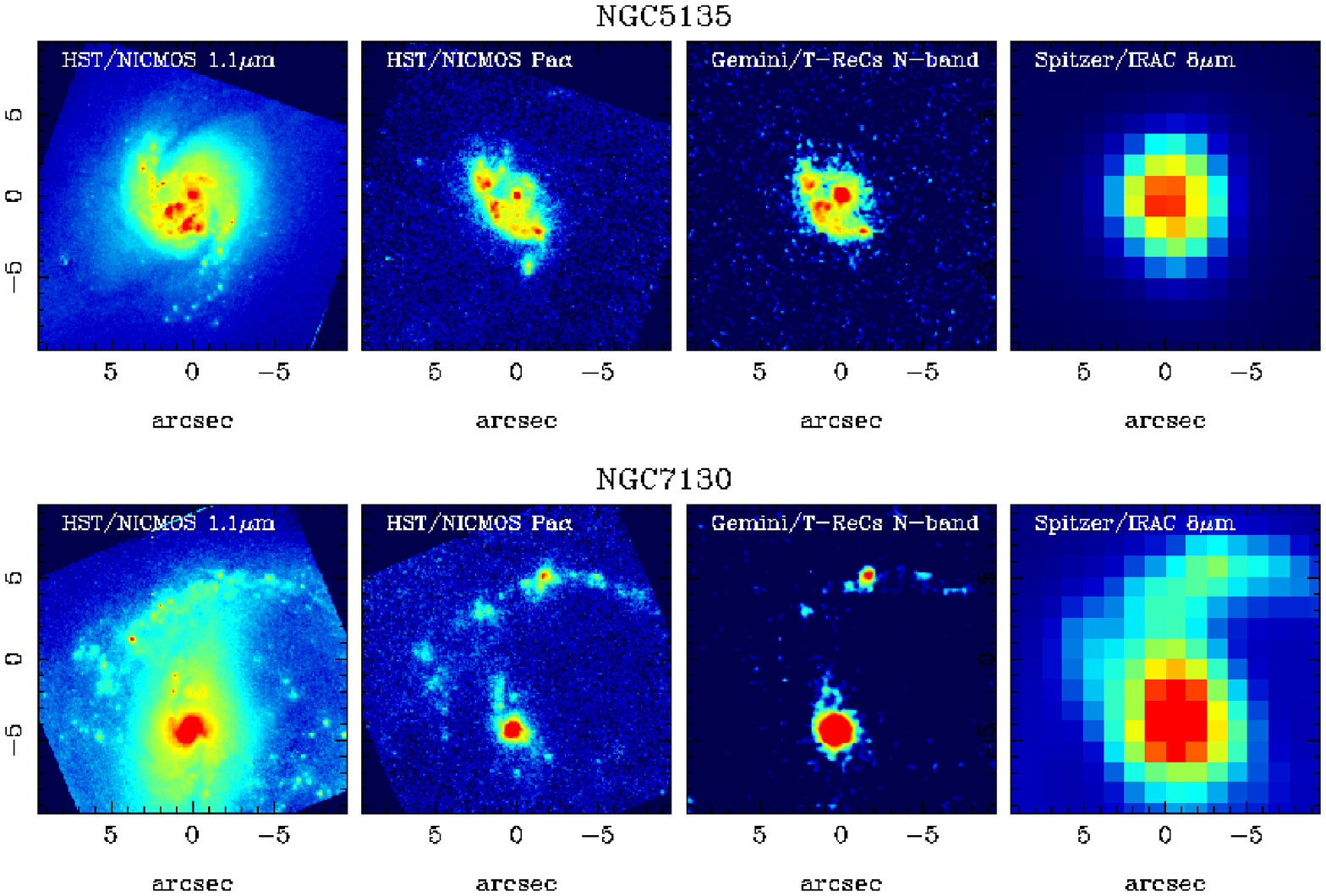}
\caption{ From left to right: {\it HST}/NICMOS $1.6\,\mu$m or $1.1\,\mu$m continuum emission,
{\it HST}/NICMOS Pa$\alpha$ emission line, Gemini/T-ReCS $10.4\,\mu$m ($N$-band) or 
$8.7\,\mu$m imaging, and {\it Spitzer}/IRAC
$8\,\mu$m imaging of the central regions of three LIRGs in our sample.
Orientation is north up,
east to the left. The two brightest mid-IR and Pa$\alpha$ sources of NGC~3256 are the
north and south nuclei of this merger LIRG.
For NGC~5135 the bright mid-IR central source is 
the Seyfert 2 nucleus,
whereas the sources
 located to the east and south of the nucleus are bright H\,{\sc
ii} regions (see \S5).  }

\end{figure*}

\section{The sample of local LIRGs}
The sample of 30 LIRGs (34 galaxies) 
was  chosen to be volume-limited to a velocity range of
$v=2750-5200\,{\rm km \,s}^{-1}$  ($d = 35-75\,$Mpc, $H_0=75\,$km s$^{-1}$
Mpc$^{-1}$)
so that the Pa$\alpha$ emission line ($\lambda_{\rm rest} =
1.876\,\mu$m) could be observed with the {\it HST}/NICMOS F190N narrow-band
filter. We also obtained NICMOS continuum images at 1.1, 1.6, and
$1.87\,\mu$m . For the velocity range
used here our sample contains  $\sim 80\%$ of all LIRGs in the {\it
  IRAS} Revised Bright Galaxy Sample (Sanders et al. 2003). 
The missing galaxies are very similar to those observed, and
thus this sample is representative of the class of local LIRGs as a whole. The
sample covers an IR 
luminosity range of 
$\log L_{{\rm   IR}[8-1000\,\mu{\rm m}]}=11-11.9\,[{\rm L}_\odot]$,
and includes a variety of morphologies and environments 
(isolated systems, galaxies in groups, 
and interacting galaxies), and nuclear activity 
(H\,{\sc ii}, Seyfert, and LINER). 
Details on the observations and near-IR properties of the sample
are given by Alonso-Herrero et al. (2006a, AAH06 hereafter).

\section{Observations}

We obtained T-ReCS imaging observations of 
all the  LIRGs in our sample that can be observed from Gemini-South. 
We focus here on the first results obtained
for four LIRGs (see Alonso-Herrero et al. 2006b).
Three LIRGs (NGC~5135, IC~4518W, and NGC~7130) 
were observed with the $N$ broad-band filter  
(central wavelength $\lambda_{\rm c}=10.36\,\mu$m) and 
NGC~3256 with the Si-2  narrow-band filter 
($\lambda_{\rm c}=8.74\,\mu$m). The 
T-ReCS observations were effectively diffraction limited (FWHM$\sim 0.30\arcsec$)  implying 
spatial resolutions of $\sim 50-100\,$pc for our LIRGs.

In addition we have retrieved archival {\it Spitzer}/IRAC $8\,\mu$m and 
MIPS $24\,\mu$m images for our sample, with spatial resolutions of
  $\sim 2\arcsec$ \, and $\sim 5\arcsec$, respectively. The {\it HST}/NICMOS, T-ReCS,
  and {\it Spitzer}/IRAC  images of three LIRGs are presented in Fig.~1.

\section{Mid-IR Emission on scales of tens-hundreds parsecs}

\subsection{Morphology}
The morphological
resemblance (see Fig.~1) between the bright mid-IR regions (traced by T-ReCS) 
and the H\,{\sc ii} regions (traced by the NICMOS Pa$\alpha$ emission) 
on scales of tens-hundreds of 
pc in LIRGs (see also Soifer et al. 2001) is remarkable.
The IRAC 8$\,\mu$m images, taking into account their limited spatial
resolution, are similar to the T-ReCs images, with the advantage that they  
are more sensitive to the diffuse emission not directly associated with the
bright H\,{\sc ii} regions (see \S5 for a detailed discussion). 
The near-IR  continuum  emission is more extended than the mid-IR
emission, 
and generally does not resemble the morphology of the ionized gas (see AAH06). 
As $A_{\rm 1.6\,\mu{\rm m}} \sim 3.3 \times A_N$, the lack of correspondence between 
the near and mid-IR (and Pa$\alpha$) emitting
regions is probably not a differential extinction effect. Rather, this 
may reflect age differences since the near-IR  continuum   
is produced mostly by stellar populations older than the ionizing stellar
populations (see e.g., Alonso-Herrero et al. 2002),  
even though highly obscured regions  tend to be associated with the 
youngest regions in LIRGs (Soifer et al. 2001). 

The nuclei are the brightest mid-IR sources in all the
galaxies. The Sy2 nuclei of 
NGC~5135 and IC~4518W appear unresolved in the T-ReCs images, in agreement
with the detection of nuclear point sources 
 in the higher spatial resolution ($\sim 0.15\arcsec$) NICMOS images (see AAH06). 
The nucleus of NGC~7130 (classified as a Sy or LINER)  
appears marginally resolved in the T-ReCS images, and the NICMOS
continuum images reveal the presence of at least three sources of similar
flux in the
central $\sim 190\,$pc. This suggests that the putative AGN in NGC~7130 
does not dominate (see \S4.2) the $N$-band nuclear emission. 
NGC~3256 is a well-studied merger galaxy, with two bright nuclei, 
which are also the brightest mid-IR sources
in this galaxy. Although it has been argued that this merger may contain
obscured AGN, the extended (${\rm FWHM} \sim 60\,$pc) nature of the near and
mid-IR continuum emission
of both nuclei suggests that such
AGN, if present, are not dominant in the mid-IR.

\subsection{AGN Contribution to the mid-IR emission}

Approximately 25\% of local LIRGs, including those in our sample,
 host spectroscopically confirmed AGN (e.g., Veilleux et al. 1995).
The Seyfert
 nucleus of IC~4518W appears to be dominate the mid-IR emission, whereas for NGC~5135 and
 NGC~7130 the Seyfert nuclei contribute $\sim 20\,\%$ or less of the $N$-band
emission within the central 10\arcsec. 
Three more galaxies  in the LIRG sample of AAH06, not observed with T-ReCS,
have spectroscopically confirmed Sy nuclei. For two of them, the B1 nucleus of 
the IC~694/NGC~3690 (Arp~299) system
(Garc\'{\i}a-Mar\'{\i}n
et al. 2006), and NGC~7469, the AGN contribution to the mid-IR emission is 
$\sim 30$\%  (Keto et al. 1997;  Soifer et al. 2003). 
For the third one, there is no mid-IR information. This is consistent
with the majority of AGN in our local LIRGs not dominating the mid-IR emission.

\begin{figure}
\includegraphics[angle=-90,width=1.\columnwidth]{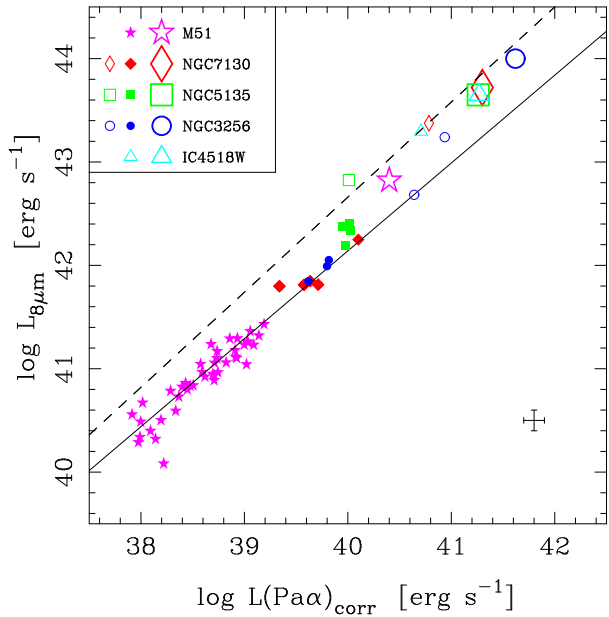}
\caption{Monochromatic ($\nu \ f_{\nu}$) 
$8\,\mu$m vs. extinction-corrected Pa$\alpha$
  luminosities.  
The small open and filled symbols are the nuclei and 
H\,{\sc ii} regions of LIRGs, respectively. 
The large open symbols for the LIRGs indicate the
  integrated properties over the NICMOS FOV ($\sim 19\arcsec \times
  19\arcsec$, between 3 and 7\,kpc for these LIRGs). The central 6\,kpc 
M51 H\,{\sc ii} regions and integrated emission from CAL05 are shown
  as star symbols. The solid line is our least-squares fit to the CAL05 M51 H\,{\sc ii} region data
  extrapolated to the LIRG luminosities. The
  dashed line is the Wu et
  al. (2005) relation for star-forming galaxies.}
\end{figure}

\section{Nuclear, H\,{\sc ii} region and integrated $8-10\,\mu$m  emission of LIRGs}

For the nearby galaxy NGC~300 Helou et al. (2004) showed that the $8\,\mu$m
emission tends to be more extended than 
the H$\alpha$ emission and highlights the rims of the
H\,{\sc ii} regions. They concluded that the $8\,\mu$m emission is more 
closely associated with photodissociation regions (PDRs) than with ionized regions.
In Fig.~2 we compare the Pa$\alpha$ and $8\,\mu$m emission for resolved H\,{\sc ii} regions,
nuclei, and integrated (central $3-7\,$kpc
emission of the four LIRGs observed with T-ReCs. 
We also show in this figure data for resolved H\,{\sc ii} knots in M51 from CAL05.
The LIRG  H\,{\sc ii} regions, although can be up to 
10 times more luminous than those in M51, tend to follow the
extrapolation of the CAL05 relation. 
Conversely, the integrated $\sim 3-7\,$kpc emission of LIRGs deviates 
significantly from the
relation found for individual H\,{\sc ii} regions, but  only  
slightly from the fit of Wu et al. (2005) for star-forming galaxies
(Fig.~3). This could arise from two causes. 
First, Wu et al. (2005) may have underestimated the reddening
(obtained from the Balmer decrements)  in these
dusty star-forming galaxies. Second, the $8\,\mu$m emission may be dominated
by diffuse emission not associated directly with the
H\,{\sc ii} regions (see below).

The LIRG nuclei (except the south nucleus of NGC~3256)  
show elevated mid-IR/Pa$\alpha$ ratios when compared to H\,{\sc ii} regions. 
Such spatial differences in the mid-IR emission of 
nuclear and circumnuclear regions are observed for our LIRGs from high-spatial
resolution T-ReCS spectroscopy (see D\'{\i}az-Santos et al. 2006). 
One possibility is  that an insufficient 
extinction correction (as nuclear 
extinctions in LIRGs tend to be higher than in extranuclear regions)   
 will produce a differential effect, making the nuclei appear more IR-luminous. 
The AGN present in three of our LIRGs may also play a role, 
as their continua are produced by dust heated to 
higher temperatures than  H\,{\sc ii} regions,  and do not 
present the strong PAH emission characteristic of H\,{\sc ii}+PDRs
(e.g., Roche et al. 2006; D\'{\i}az-Santos et al. 2006).

\begin{figure*}
\includegraphics[width=2.\columnwidth]{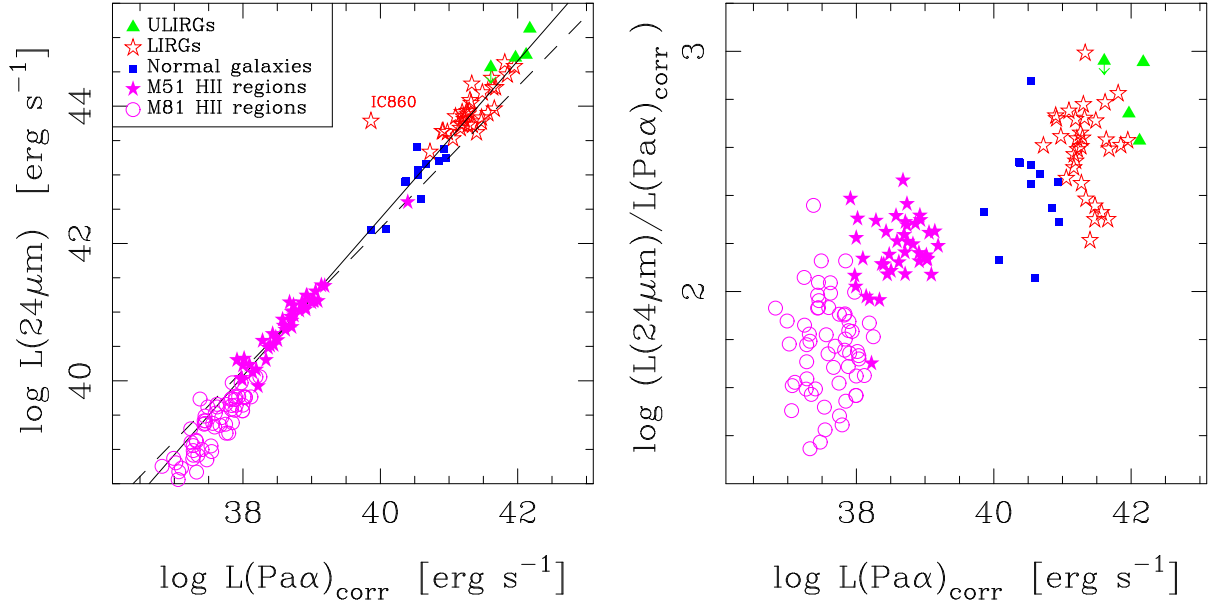}
\caption{Monochromatic $24\,\mu$m
  luminosities vs. extinction-corrected Pa$\alpha$ luminosities (left)  
for LIRGs, normal galaxies (B\"oker et al. 1999), ULIRGs (Murphy et al. 2001)
ULIRGs, and H\,{\sc ii} regions in M51 and M81 (see
  text
and AAH06).
The solid line is the fit to the M51 H\,{\sc
  ii} and the normal galaxies and LIRGs, and dashed line 
the CAL05 fit to the M51 H\,{\sc
  ii} knots. 
The $24\,\mu$m/Pa$\alpha$ ratio vs. the extinction-corrected Pa$\alpha$
  luminosity (right). Symbols are as
  in left panel.}

\end{figure*}

Since a variety of H\,{\sc ii} regions in M51 and LIRGs have a similar 
$8\,\mu$m vs. Pa$\alpha$ relation
we conclude that their $8\,\mu$m emission in the {\it Spitzer}/IRAC band 
is well characterized by a thermal continuum plus PAH features
with no strong variations over the range of conditions
probed here (e.g., metalicity near or over solar, see
AAH06 and CAL05). 

However, the integrated
central $3 - 7\,$kpc $8\,\mu$m vs. Pa$\alpha$ emission differs
significantly in all these environments from the relation found for
the individual H\,{\sc ii} regions. This may be explained by the presence, in
addition to the bright and compact 
H\,{\sc ii} regions, of a  more diffuse
and extended $8\,\mu$m 
component (see Fig.~1, and Helou et al. 2004; D\'{\i}az-Santos et
al. 2006). This extra  
emission at $8\,\mu$m  would be produced not by local,
strong  ionizing sources, but by the diffuse radiation field  that permeates
the ISM. As such, the spectrum in these
regions would be characterized by a weak continuum and strong PAH
features with a large equivalent width. This has already been confirmed for
one of the LIRGs in our sample as discussed in detail by D\'{\i}az-Santos et
al. (2006). Hence when compared to 
individual H\,{\sc ii} regions, an excess of $8\,\mu$m/Pa$\alpha$ emission can
be expected for the integrated properties over a few kpc. 
This is further supported by the fact that the
central  $3 - 7\,$kpc emission of LIRGs falls only slightly below
the H$\alpha$  vs. $8\,\mu$m relation found for the integrated
properties of the galaxies studied by Wu et al. (2005), as
indicated in Fig.~2. We conclude that calibrations of the SFR in terms of the
$8\,\mu$m emission for distant
galaxies should be based on the integrated mid-IR emission of galaxies, not
that of the H\,{\sc ii} regions alone.

\section{The $24\,\mu$m emission as a SFR tracer for dusty star-forming galaxies}

Helou et al. (2004) showed that in contrast with the $8\,\mu$m emission, the
$24\,\mu$m emission of NGC~300 appears to be more centrally peaked in
star-forming regions. 
In Fig.~3 (left) we compare the monochromatic $24\,\mu$m
luminosities and the extinction-corrected Pa$\alpha$ luminosities (see AAH06)
for all the LIRGs in our sample. We also include a small comparison sample of
normal galaxies and ULIRGs, and data for resolved star-forming regions 
in M51 from CAL05, and in M81 from PGPG06.  

The integrated emission of LIRGs, ULIRGs, and normal galaxies
seems to continue the relation between the extinction-corrected Pa$\alpha$
luminosity and the $24\,\mu$m luminosity observed by CAL05 
for the M51 central H\,{\sc ii} knots, unlike the case of the $8\,\mu$m emission. 
The fit to the LIRGs, ULIRGs, normal galaxies, and M51 H\,{\sc ii} knots
provides a slope of $1.148\pm0.013$ (Fig.~3; Wu et al. 2005; 
AAH06), only slightly larger
than the slope of the CAL05 fit to the M51 H\,{\sc ii} knots. The M81 H\,{\sc ii}
regions, on the other hand, 
appear to be offset with respect to M51 H\,{\sc ii} regions, normal
galaxies,  and LIRGs. This
behavior could arise from a lower UV absorption efficiency in the relatively
low luminosity and lightly obscured H\,{\sc ii} regions in M81, as well as to
uncertain extinction corrections (see discussion by PGPG06).
There are galaxy-to-galaxy variations in the mid-IR to
Pa$\alpha$ ratios (Fig.~3, right), as also noted by CAL05. The variation of 
the $L(24\,\mu{\rm m})/L({\rm Pa}\alpha)_{\rm corr}$ ratio
with the Pa$\alpha$ luminosity (i.e., SFR) 
between the M51 H\,{\sc ii} knots and our LIRGs and  ULIRGs is similar to the   
$L_{24\,\mu{\rm m}}$/SFR ratios found by CAL05. 
In fact, the gap between the M51 H\,{\sc ii}
knots and the LIRGs/ULIRGs seen in our figure is populated by their UV-selected
starbursts.

%In the next two sections we discuss two possible causes 
%for the deviation from strict proportionality of the $L(24\,\mu{\rm m})$ vs. 
%$L({\rm Pa}\alpha)_{\rm corr}$ relation.

There are two possibilities which might explain the deviation from strict
proportionally of the $24\,\mu$m vs. Pa$\alpha$ relation. First 
the nuclear extinctions of  LIRGs may be underestimated (the values averaged
over  the Pa$\alpha$ emitting regions are $A_V \sim 2-6\,$mag, see AAH06),
whereas some LIRGs and ULIRGs are known to contain highly obscured regions, 
usually associated with the nuclei of the galaxies (e.g., Genzel et
1998; Alonso-Herrero et al. 2000). From  the observed ratios of mid-IR
emission lines of star-forming galaxies there is evidence that some of the
  youngest stars in LIRGs may still embedded in high density 
ultracompact H\,{\sc ii} 
  regions and hidden from us by large amounts of extinction (Rigby \& Rieke
  2004). In such regions the dust competes
increasingly effectively for ionizing photons and UV continuum
photons, so that an increasing fraction of the luminosity is expected 
to emerge in the IR. The second possibility would be if the LIRG emission is
dominated by AGN, but as discussed in \S4.2  AGN do not appear to dominate the mid-IR emission in our sample
of LIRGs.

\subsection{The empirical calibration of SFR vs. $24\,\mu$m}
In deriving his relationship between the SFR and the IR luminosity Kennicutt (1998) 
assumed
that the great majority of the luminosity from young stars would be
absorbed by dust and reradiated in the far-IR.  
Although this assumption is likely to be correct,
observationally there may be other contributions to the total IR
luminosity from older populations of stars or other luminosity sources. 
Therefore, a true total IR
luminosity measurement may overestimate the recent star formation in a
galaxy, although estimates based on {\it IRAS} measurements alone are less
subject to these issues because they only poorly sample the output of the
cold dust (because the longest band is at 100$\mu$m). Second, Kennicutt
assumed a direct proportionality between the H$\alpha$ luminosity and the
total IR. We find empirically that the increasing absorption
efficiency in more luminous and obscured galaxies leads to a
deviation from strict proportionality,  causing the IR output to rise 
with increasing star-forming luminosity.

Assuming Case B recombination, and using  
the relation between SFR and H$\alpha$ quoted by Kennicutt (1998)
and the fit between the $24\,\mu$m and Pa$\alpha$ luminosities (AAH06)
we derive the following  relation between the SFR rate and the $24\,\mu$m
luminosity for luminous, dusty galaxies: 

\begin{equation}
{\rm SFR} = 8.45 \times 10^{-38}\, (L(24\,\mu{\rm m}))^{0.871}    
\end{equation}

\noindent where the SFR is in ${\rm M}_\odot\,{\rm yr}^{-1}$ 
and the $24\,\mu$m luminosity is in 
${\rm erg\,s}^{-1}$. 
This relation is analogous to the widely used relation between SFR and IR
luminosity (Kennicutt 1998) but it is not affected by the uncertain
contribution to the total IR luminosity of a galaxy of 
dust heating from old
stars.

Summarizing, we have used the sensitivity of {\it Spitzer} and the high
spatial resolution of T-ReCS on the Gemini-South telescope to study the mid-IR
properties of LIRGs. The T-ReCS $8-10\,\mu$m imaging observations of LIRGs have allowed us
to resolve the nuclear emission (star formation and/or AGN) from that of
circumnuclear H\,{\sc ii} regions. The comparison of the
$8\,\mu$m/Pa$\alpha$ ratios of the integrated and resolved H\,{\sc ii} regions
of LIRGs reveals an $8\,\mu$m 
diffuse component, not directly related to the ionizing stars,
that can be as luminous as that from the resolved H\,{\sc ii} regions.  
Therefore the calibration of the SFR for distant 
galaxies should be based on the 
integrated mid-IR emission of nearby galaxies, not that of H\,{\sc ii} regions
alone. We provide a calibration of the SFR in terms of the integrated 
$24\,\mu$m luminosity. Similar mid-IR studies  with the
GTC/CanariCam system will be possible in the near future.

%\section{}   %%% Top level section head (remove "%" symbol)
%\subsection{}   %%% Second level section head (remove "%" symbol)
%\subsubsection{}   %%% Lowest level section head (remove "%" symbol)
%\section*{}	%%% Unnumbered top level section head (remove "%" symbol)
%\subsection*{}   %%% Unnumbered second level section head (remove "%" symbol)

\acknowledgements %%% Text of acknowledgements runs on after this command.

I thank L. Colina, T.
D\'{\i}az-Santos, C. Packham, P. P\'erez-Gonz\'alez, J. Radomski, 
G. Rieke, M. Rieke, S. Ryder,  and C. Telesco
for their help. 

Support was provided by 
the Spanish PNE (ESP2005-01480) and the NSF (0206617).
Based on observations obtained at the Gemini Observatory, which is operated by 
AURA, Inc., under a cooperative agreement
with the NSF on behalf of the Gemini partnership:  NSF (USA),  
PPARC (UK), NRC (Canada), CONICYT (Chile), ARC
(Australia), CNPq (Brazil) and CONICET (Argentina).
Based on observations with the NASA/ESA {\it HST}, 
obtained at the STScI, which is operated by AURA, Inc., under NASA contract
NAS 5-26555.

%%% THE BIBLIOGRAPHY
%%%
%%% CONSULT SECTION 3 OF "INSTRUCTIONS FOR AUTHORS" FOR HOW TO USE NATBIB.
%%% AUTHORS ARE ENCOURAGED TO USE EITHER THE "THEBIBLIOGRAPY" ENVIRONMENT
%%% BY UNCOMMENTING (DELETING THE "%" SYMBOL) THE COMMANDS BELOW, OR BY
%%% USING THE BIBTEX ENVIRONMENT. TO FIND OUT WHICH IS APPLICABLE TO YOUR
%%% CONTRIBUTION, CONSULT THE VOLUME EDITORS FOR YOUR PROCEEDINGS.
%%%

\end{document}